\documentclass[nohyperref]{article}

\usepackage{microtype}
\usepackage[pdftex]{graphicx}
\usepackage{booktabs} 
\usepackage{multirow}

\usepackage{hyperref}
\usepackage{url}

\usepackage[]{caption, subcaption}

\usepackage{color,soul} 

\usepackage[accepted]{icml2022}

\usepackage{amsmath}
\usepackage{amssymb}
\usepackage{mathtools}
\usepackage{amsthm}

\usepackage[capitalize,noabbrev]{cleveref}

\theoremstyle{plain}

\theoremstyle{definition}

\theoremstyle{remark}

\usepackage[textsize=tiny]{todonotes}

\icmltitlerunning{Robustness Evaluation of Deep Unsupervised Learning Algorithms}

\begin{document}

\twocolumn[
\icmltitle{Robustness Evaluation of Deep Unsupervised Learning Algorithms for Intrusion Detection Systems}



\icmlsetsymbol{equal}{*}

\begin{icmlauthorlist}
\icmlauthor{D'Jeff K. Nkashama}{udes}
\icmlauthor{Arian Soltani}{udes}
\icmlauthor{Jean-Charles Verdier}{udes}
\icmlauthor{Marc Frappier}{udes}
\icmlauthor{Pierre-Martin Tardif}{udes}
\icmlauthor{Froduald Kabanza}{udes}
\end{icmlauthorlist}

\icmlaffiliation{udes}{GRIC, Université de Sherbrooke, Sherbrooke, QC, Canada}

\icmlcorrespondingauthor{D'Jeff K. Nkashama\\}{djeff.nkashama.kanda@usherbrooke.ca}




\icmlkeywords{Anomaly detection,Unsupervised learning, Robustness,ICML}

\vskip 0.3in
]



\printAffiliationsAndNotice{}  

\begin{abstract}

Recently, advances in deep learning have been observed in various fields, including computer vision, natural language processing, and cybersecurity. Machine learning (ML) has demonstrated its ability as a potential tool for anomaly detection-based intrusion detection systems to build secure computer networks. Increasingly, ML approaches are widely adopted than heuristic approaches for cybersecurity because they learn directly from data. Data is critical for the development of ML systems, and becomes potential targets for attackers. Basically, data poisoning or contamination is one of the most common techniques used to fool ML models through data. This paper evaluates the robustness of six recent deep learning algorithms for intrusion detection on contaminated data. Our experiments suggest that the state-of-the-art algorithms used in this study are sensitive to data contamination and reveal the importance of self-defense against data perturbation when developing novel models, especially for intrusion detection systems.

\end{abstract}

\section{Introduction}
Nowadays, information technology has become a central part of our daily activities, including communication, travel, manufacturing, banking, education, and work. While these technologies are beneficial to users, they also allow adversaries to conduct malicious activities on cyberinfrastructures. These malicious activities include but are not limited to information stealing, denial of service, manipulation of cyberinfrastructure components, and data poisoning.  Cyber threats spare no one, from individuals, businesses, banks to governments. Therefore, cyberinfrastructures must be protected against these permanent and harmful cyber threats; firewalls, spam filters, anti-virus, and intrusion detection systems (IDS) form the ecosystem of network defense \cite{dua2016data}.


Recent advances in machine learning (ML) play a pivotal role in cybersecurity and contribute to the progress of this field at a fast pace. Many ML techniques on anomaly detection (AD) have found their applications in cybersecurity. An anomaly is an observation that considerably deviates from  what is deemed normal observations \cite{alvarez2022revealing}. Depending on the situation, such an observation is considered unusual, irregular, atypical, inconsistent, unexpected, rare, erroneous, faulty, fraudulent, malicious, etc~\cite{ruff2021unifying, chalapathy2019deep}.

AD approaches can be categorized as follows: probabilistic methods such as DAGMM~\cite{zong2018deep}, which estimate data probability density function and predict samples laying in the low-density region as anomalies; reconstruction based methods such as MemAE~\cite{gong2019memorizing} and DUAD~\cite{li2021deep}, which assume that normal data is compressible, and flag as anomalous, data samples that can not be reconstructed from their compression; distance based models such as LOF~\cite{breunig2000lof}, which predict anomalies by using their distances from normal data; one-class classification approaches such as OC-SVM~\cite{scholkopf1999support}. Examples of AD applications in cybersecurity include malware detection, spam filtering, and intrusion detection systems~\cite{crawford2015survey, liu2020anomaly}.

Training ML models for AD requires a nearly large amount of data usually collected from various sources, including sensors, actuators, processes, network traffic, and people. As the volume of data grows, it becomes more and more susceptible to contamination. One source of contamination might be an ongoing attack campaign, yet unnoticed, during data collection. Moreover, adversaries are aware that data is crucial to developing ML systems for intrusion detection, so they attempt to attack it by injecting malicious samples into the training set \cite{hayase2021spectre} -- this is known as data poisoning or data contamination. Adversaries' goal is to hamper model performance and probably create a backdoor for subsequent intrusions.  For example, in spam filtering, an attacker will attempt to evade the spam detection system by misspelling words most likely to appear in spam and adding words likely to occur in legitimate emails. An attacker will slightly modify the malware code without altering its functionality while making it undetectable in malware detection.


Currently, AD training pipelines operate under the assumption that the training data is clean. However, this is not always verified because checking data cleanliness is expensive for large-scale datasets \cite{hayase2021spectre}. Therefore, it is crucial for AD models to be robust against contamination during training, especially in the use case of intrusion detection. Although the robustness of AD models has already been studied, the most recent AD models are rarely exposed to and tested on contaminated data. At the same time, that could be a decisive factor in their real-world applications.

In this paper, we assess the robustness of six state-of-the-art deep learning models for AD, with different levels of training set contamination following our evaluation protocol. These models are: ALAD \cite{zenati2018adversarially}, Deep auto-encoder \cite{chen2018autoencoder}, DAGMM \cite{zong2018deep}, DSEBM \cite{zhai2016deep}, DUAD \cite{li2021deep}, and NeuTraLAD \cite{qiu2021neural}. We employ CIC-CSE-IDS2018 dataset and two widely used benchmark datasets: KDDCUP and NSL-KDD\footnote{\href{https://www.unb.ca/cic/datasets/nsl.html}{https://www.unb.ca/cic/datasets/nsl.html}}. These are all cybersecurity datasets simulating computer networks.

The rest of the paper is organized as follows: In Section~\ref{relwork}, we describe the related work. Section~\ref{evprot} presents our evaluation protocol. We discuss different experimental results in Section~\ref{experiements}. In Section~\ref{conclu}, we conclude the paper and discuss future works.

\section{Related Work}
\label{relwork}

Attacking ML systems through data manipulation can be done at training time or testing time. Evasion attacks and data poisoning (also known as data contamination) are among the most used techniques by adversaries to fool ML systems~\cite{biggio2018wild,ilyas2019adversarial}. While evasion attacks consist of manipulating data at test time, data poisoning is performed at training time. In this work, we focus on data manipulation during training.

\subsection{Data poisoning attacks}
Attacks that insert poisoned instances into the training data are known as data poisoning. These attacks aim to increase the classification errors at test time. The adversary's goal can be to increase misclassification generically, i.e., the classes where they occur do not matter. Alternatively, the adversary may target a specific class where the misclassification should occur.

Research on data poisoning has gradually surged and proposed sophisticated techniques. For instance, earlier works on SVMs~\cite{biggio2012poisoning}, feature selection~\cite{xiao2015feature}, and PCA~\cite{rubinstein2009antidote}, as well as initial efforts with deep learning that mostly focused on lowering the model quality~\cite{munoz2017towards}. Later, this area was dominated by research based on generative models~\cite{goodfellow2014generative}. With generative models, attacks could aim toward creating a backdoor, where samples with specific and easy to manipulate characteristics would bypass detection by the model~\cite{wang2020certifying}. We note that less attention has been paid to data poisoning in the case of intrusion detection systems. Our work studies data poisoning with both generic and specific types of attacks. 

\subsection{Deep Unsupervised Anomaly Detection}
Anomaly detection in an unsupervised setting is challenging since the absence of explicit labels limits exploitable information. Classical approaches to unsupervised AD work well with lower dimensions. However, as the feature space grows, they suffer from the curse of dimensionality \cite{li2021deep}. Deep learning architectures, including autoencoders, have shown significant potential to mitigate this issue.

Unsupervised AD consists of three approaches.
Reconstruction-based methods operate under the assumption that anomalies are harder to regenerate after compression. MemAE \cite{gong2019memorizing} augments encoder-decoder reconstruction with a memory module to store common patterns. DUAD \cite{li2021deep} assumes normal sample clusters have lower variance and iteratively refines the clusters by their encoded representation to exclude and alienate anomalies. 
Reconstruction-based methods have also been implemented with GANs \cite{zenati2018adversarially}.
Probabilistic and density-based methods exploit the isolation of anomalous samples and recognize them by lower density of samples in their feature space. As an example, DAGMM \cite{zong2018deep} is one the recently proposed methods of this approach, utilizing reconstruction error and latent representation to determine the parameters of a Gaussian Mixture Model (GMM), which in turn appoints anomalous samples by their log-likelihood. 
One-class classifiers are another anomaly detection approach that learns the boundary enveloping normal instances.  Few prior works have been proposed this approach in an unsupervised context~\cite{ruff2018deep}. Recent works employ self-supervised pre-trained feature extractors for one class classification \cite{bergman2020classification, golan2018deep, sohn2020learning}. 
Even though a lot of progress has been made using deep learning architectures, we aim to measure the newly proposed models from the perspective of robustness to training data contamination.

\subsection{Robustness in Anomaly Detection}
The robustness of a model is defined as its resistance to exposure to unexpected data for the same task. In real-world applications, data contamination and adversarial attacks threaten the capabilities of anomaly detection models. 
Robustness to noise contamination was explored in SVMs \cite{xu2009robustness}. This concept has also been introduced in principal component analysis under the name of Robust PCA (RPCA)~\cite{zhou2017anomaly}. RPCA utilized low-rank matrix decomposition to achieve robustness, and the same has been applied to extend Autoencoder~\cite{goodge2020robustness}.

To the best of our knowledge, the problem of robustness against noise contamination has not been clearly and carefully studied in the cybersecurity field. We believe that this problem represents a ubiquitous concern in the intrusion detection task. This work investigates the robustness of recent intrusion detection methods on different datasets, including a recent dataset that simulates a complex network. Our work will better inform the selection of an intrusion detection model whose quality should include resilience to data perturbations at training time.




\section{Evaluation Protocol}
\label{evprot}
This section presents the data split strategy, the metrics, and the choice of the decision threshold.
\begin{table*}[!tbp]
    \caption{Datasets statistics.}
    \label{dataset-info}
    \vskip 0.15in
    \centering
    \begin{tabular}{l|c|c|c}
        \hline
         Dataset &  \# Samples & \# features & Attacks ratio\\
         \hline
         KDDCUP 10\% & 494 021 & 42 & 0.1969\\
         NSL-KDD & 148 517 & 42 & 0.4811\\
         CSE-CIC-IDS2018 & 16 232 944 & 83 & 0.1693\\
         \hline
    \end{tabular}
\end{table*}

\textbf{Data split.} Following \citet{zong2018deep}, the training set contains $50\%$ of normal data samples randomly drawn, but we add a ratio $c \in \{0, 5, 8, 12\}$ percent of attack data. The test set contains $50\%$ of the rest of the normal data plus a percentage $(1-p)$ of attack data with $p \in \{20, 40\}$. The remaining $p$ percent of attack data is in a set that we call the contamination set. More specifically, the ratio $c$ of attack data utilized to contaminate the training set stems from the contamination set. The rationale behind this data split strategy is to preserve the consistency over different sets of experiments and fairness in evaluating different algorithms, as pointed out by \citet{alvarez2022revealing}. Keeping constant the proportion of normal data and attack data in the test set allows for a good assessment of the impact of different levels of training set contamination on models’ performances. To draw reliable conclusions from our experiments, we furthermore apply this data split strategy in each run to capture the variance due to data sampling in addition to the variance due to parameters initialization as motioned in \cite{bouthillier2021accounting}.

\textbf{Class of interest.} Since cybersecurity datasets have a significant class imbalance, we consider the attack data as the class of interest. Attack data is often the minority class in a real-world application. Due to this scarcity, it is important to evaluate how good models perform on it. 

\textbf{Performance metrics and threshold.} We report the average and standard deviation of F1-score, precision, and recall over 20 runs for every ratio $c$ of the training set contamination. All metrics are computed on the attack data class, which is considered as the class of interest. These metrics are suitable for problems with imbalanced class datasets~\cite{chicco2020advantages}. However, F1-score, precision, and recall calculation require setting a decision threshold on the samples' scores. In our experiments, we use 20\% of the test set to find the decision threshold and then test it on the remaining 80\%. We choose the decision threshold such that the F1-score -- the harmonic mean of precision and recall -- is the best that the model could achieve on a specific dataset.


\section{Experiments}
\label{experiements}

We present datasets and models' descriptions and discuss experimental results.

\subsection{Datasets}
\begin{figure*}[!tbp]
     \centering
     \begin{subfigure}[b]{0.45\textwidth}
         \centering
         \includegraphics[width=\linewidth]{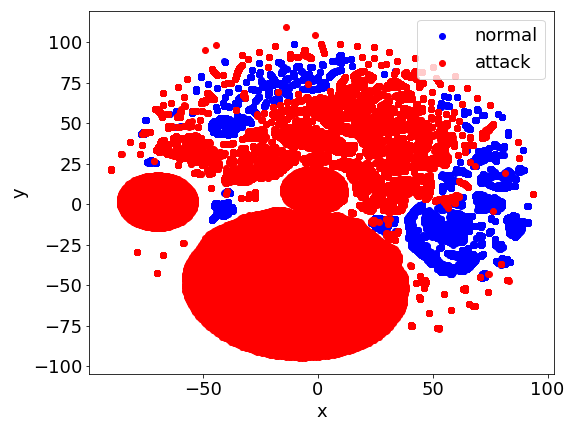}
         \caption{}
         \label{fig:tsnekdd}
     \end{subfigure}
     \hfill
     \begin{subfigure}[b]{0.45\textwidth}
         \centering
         \includegraphics[width=\linewidth]{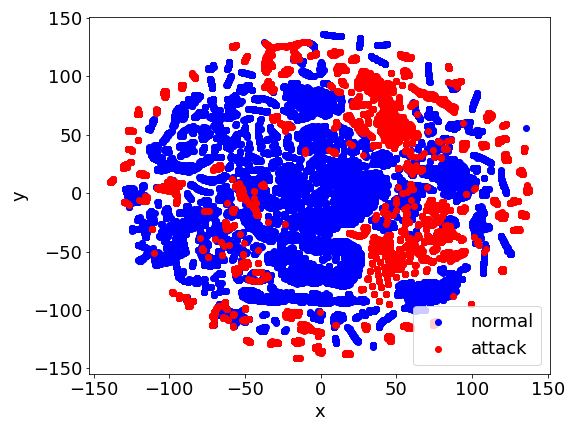}
         \caption{}
         \label{fig:tsnensl}
     \end{subfigure}
     \hfill
     \begin{subfigure}[b]{0.45\textwidth}
         \centering
         \includegraphics[width=\linewidth]{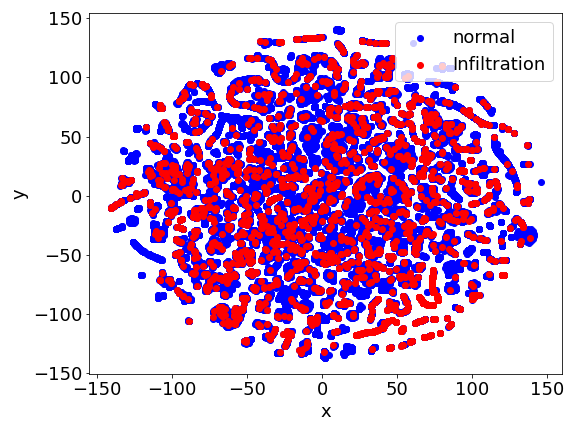}
         \caption{}
         \label{fig:tsnecic}
     \end{subfigure}
        \caption{Visualization of normal traffic and attack data from the three datasets in a two-dimensional space by t-SNE \cite{van2008visualizing}: (a) and (b) Display normal and attack data from KDDCUP and NSL-KDD datasets, respectively, with all types of attacks combined into one class.   (c) Displays normal data and \textit{infiltration} attack data from the CSE-CIC-IDS2018 data set.}
        \label{fig:tnseviz}
\end{figure*}
\begin{figure*}[!tbp]
     \centering
     \begin{subfigure}[b]{0.33\textwidth}
         \centering
         \includegraphics[width=\linewidth]{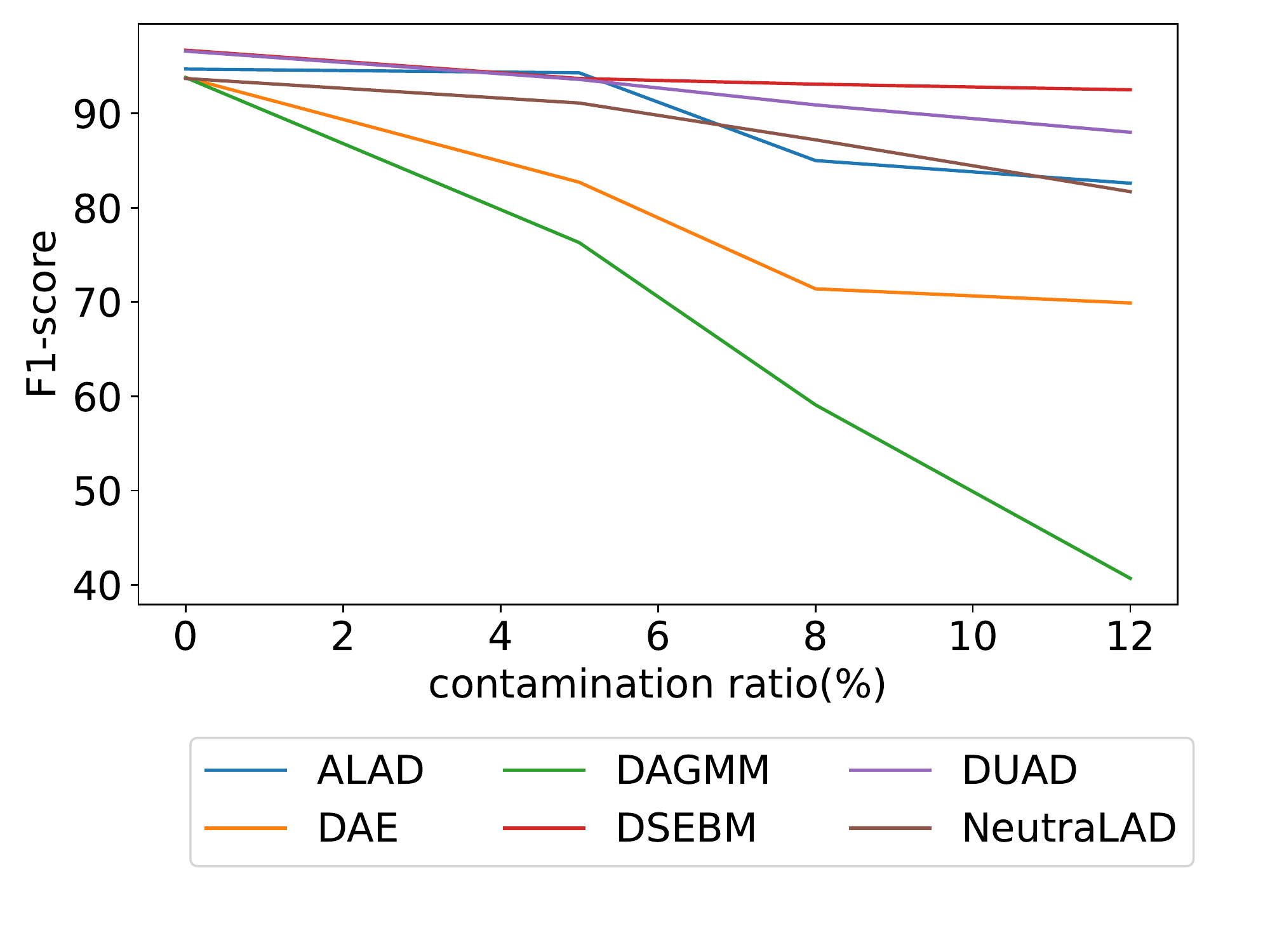}
         \caption{F1-scores on KDDCUP dataset}
         \label{fig:f1kdd}
     \end{subfigure}
     \hfill
     \begin{subfigure}[b]{0.33\textwidth}
         \centering
         \includegraphics[width=\linewidth]{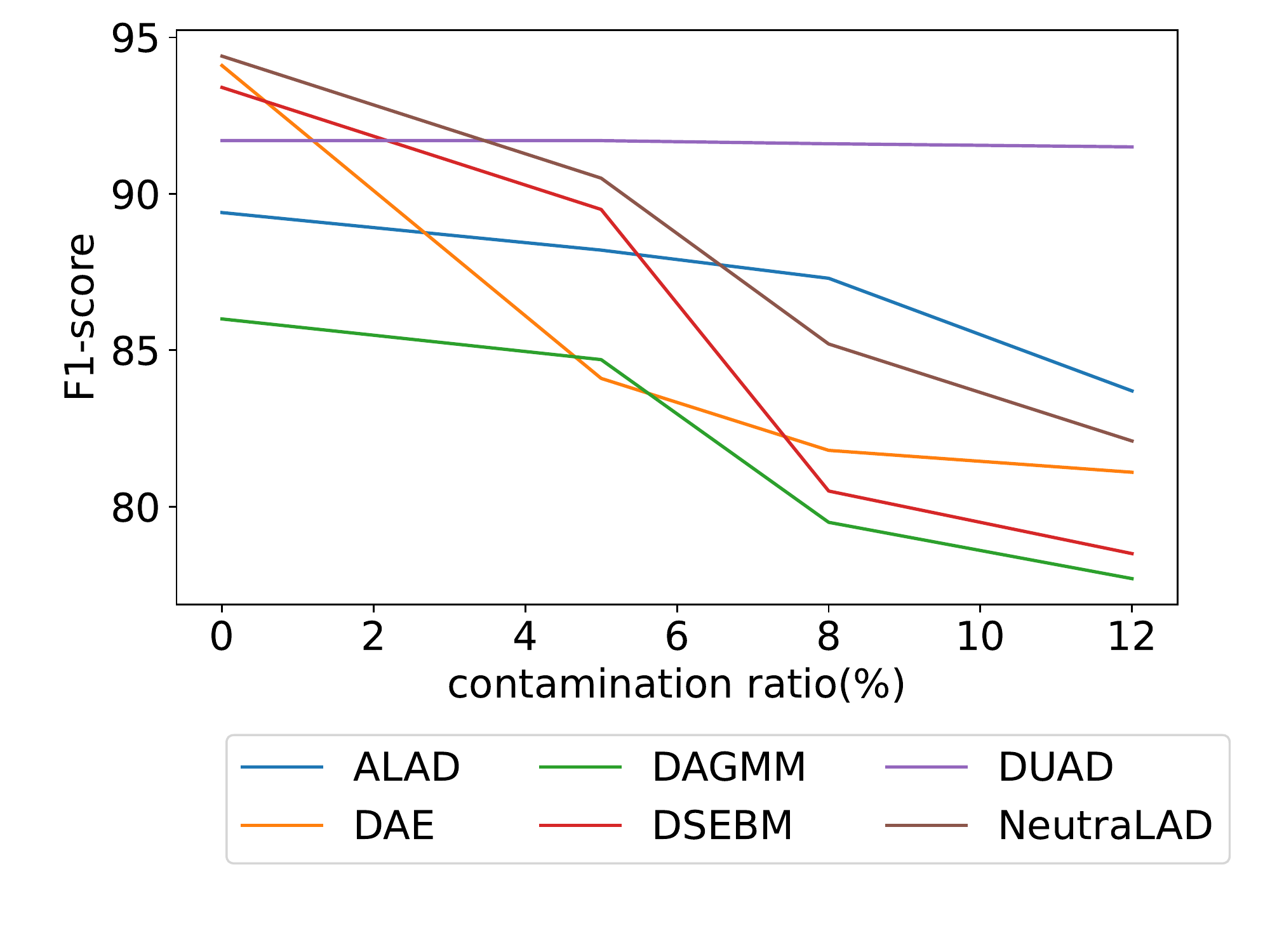}
         \caption{F1-scores on NSL-KDD dataset}
         \label{fig:f1nsl}
     \end{subfigure}
     \hfill
     \begin{subfigure}[b]{0.33\textwidth}
         \centering
         \includegraphics[width=\linewidth]{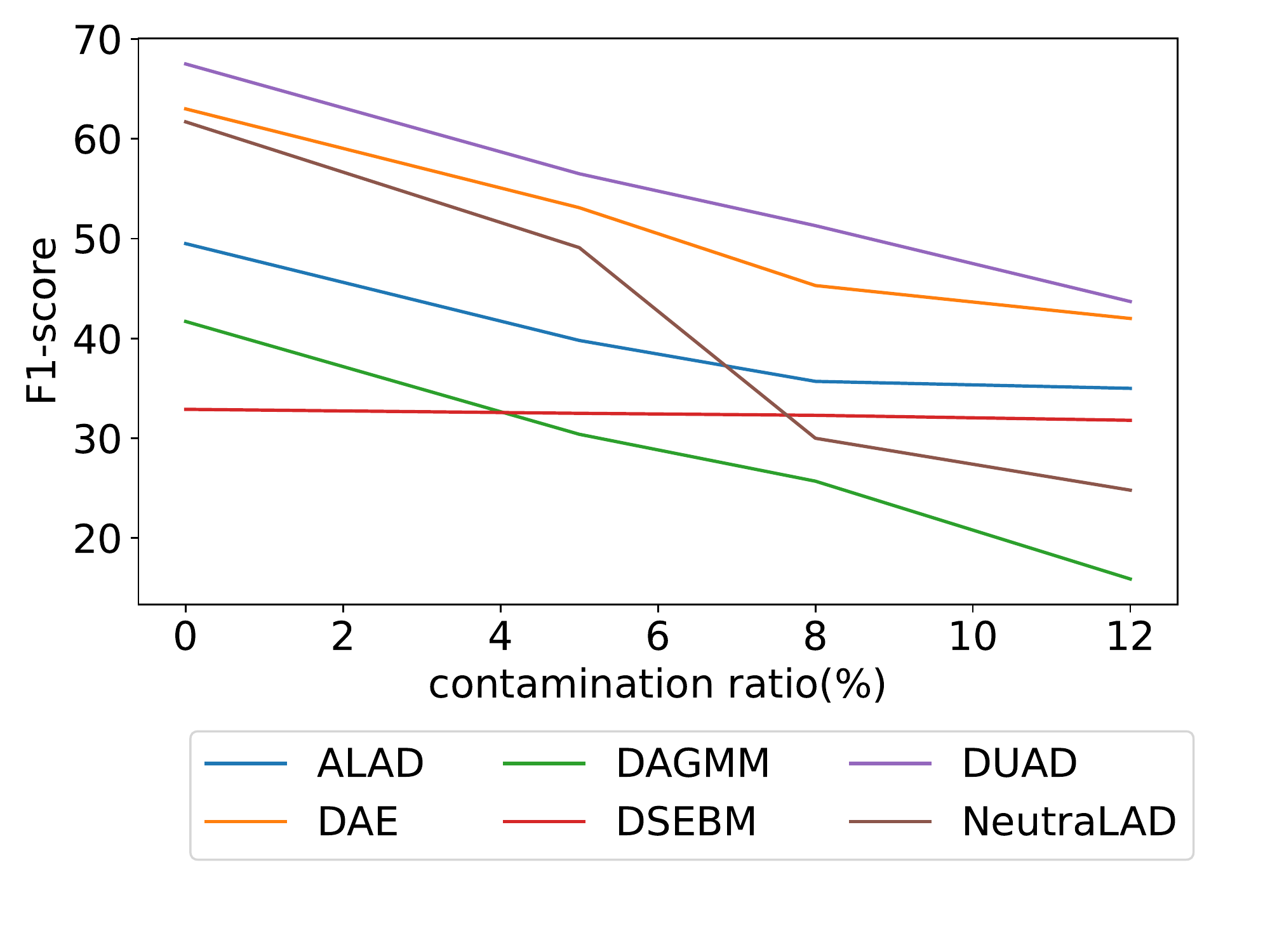}
         \caption{F1-scores on CSE-CIC-IDS2018 dataset}
         \label{fig:f1cic}
     \end{subfigure}
        \caption{Visualization of F1-scores of the six models on all the datasets, with the training data contamination ratio on the x-axis.}
        \label{fig:tnseviz}
\end{figure*}

\begin{itemize}
    \item \textbf{KDDCUP}\footnote{\href{http://kdd.ics.uci.edu/databases/kddcup99/kddcup99.html}{http://kdd.ics.uci.edu/databases/kddcup99/kddcup99.html}} is a cyber-intrusion detection dataset that is 20 years old, and has some issues such as duplicated samples~\cite{tavallaee2009detailed},  but it is still widely used as a benchmark dataset. It contains data related to normal traffic and attacks simulated in a military network environment. The Attacks include DoS, R2L, U2R, and probing. The dataset has 41 features, with 34 continuous, and the rest are categorical features. It counts 80\% of attack data and 20\% of normal data, unlike what one could expect. So, we swapped data labels to reflect a real-world scenario where the majority class is the class of normal data. We use the version that holds only 10 percent of the original data.
    
    
    \item \textbf{NSL-KDD} is a revised version of the KDDCUP dataset provided by the Canadian Institute of Cybersecurity (CIC). Some inherent issues, such as duplicated samples from the original KDD dataset, are solved in this version. Further details on this dataset are discussed in \cite{tavallaee2009detailed}.

    \item \textbf{CIC-CSE-IDS2018} is also provided by CIC in collaboration with the Communication Security Establishment. Unlike the KDDCUP dataset, this dataset is recent, and it contains normal traffic plus attacks data simulated on a complex network. Attack types include Brute-force, Heartbleed, Botnet, DoS, DDoS, Web attacks, and infiltration of the network from the inside~\cite{sharafaldin2018toward}.

\end{itemize}

Importantly, we combined all types of attacks data samples into one class, i.e., \emph{``attack''} for all datasets. Especially for the CSE-CIC-IDS2018 dataset, we kept the original labels to investigate how the contamination of training data with some filtered types of attacks affects each class. Table \ref{dataset-info} shows the statistics of the three datasets. We applied Min-Max scaling to continuous features and one-hot encoding to categorical features for the three datasets.

\subsection{Models}

This section briefly presents the six state-of-the-art anomaly detection models implemented for this study. Our code is available on Github\footnote{\href{https://github.com/intrudetection/robevalanodetect}{https://github.com/intrudetection/robevalanodetect}}.


\textbf{Deep Auto-Encoder \cite{chen2018autoencoder}}. DAE is a neural network with two components; an encoder whose output is the input's low-dimensional representation (latent representation) and a decoder accounting for reconstructing the input from its low-dimensional representation. For the loss function, we have added an $L_2$ regularization on the latent variable in addition to the standard reconstruction error. The reconstruction error is also used as the anomaly score.

\textbf{Deep Unsupervised Anomaly Detection \cite{li2021deep}}. DUAD is a method that uses a DAE for anomaly detection. It is based on the hypothesis that anomalies in the training set are approximately samples with high variance distribution. Unlike DAE, it applies a distribution clustering to select a subset of normal data from the training set after a fixed number of iterations. The reconstruction error is used as the anomaly score.

\textbf{Deep Structured Energy Based Models for Anomaly Detection \cite{zhai2016deep}}. DSEBM, as its name suggests, is an energy-based model. It learns the energy function of input data through a neural network with structure. The algorithm provides two scoring functions based on energy and another based on the reconstruction error. For our experiments, we consider the energy-based anomaly scoring function. With the energy scoring function, samples with high energy are classified as anomalies and those with low energy are classified as normal.

\textbf{Deep Auto Encoding Gaussian Mixture Model \cite{zong2018deep}}. DAGMM consists of two neural networks: an auto-encoder and an estimation network. These networks are trained in an end-to-end fashion. The model concatenates the output of the encoder -- the latent representation of the input -- and the reconstruction error, then feeds them to the estimation network, whose output is used to compute the parameters of a Gaussian Mixture Model (GMM). Specifically, the estimation network is used to obtain sample likelihood, which is also considered the anomaly score. 




\textbf{Adversarially Learned Anomaly Detection \cite{zenati2018adversarially}}. ALAD extends BiGAN~\cite{donahue2016adversarial} by adding two more discriminators to ensure data-space and latent-space cycle consistencies. The scoring function is the reconstruction error on the output of an intermediate layer of one of the discriminators.

\textbf{Neural Transformation Learning for Deep Anomaly Detection Beyond Images \cite{qiu2021neural}}. NeuTraLAD is a self-supervised learning method for anomaly detection. It combines contrastive learning with the idea of learning data transformation through neural networks to do data augmentation with tabular data. The loss function is defined to maximize agreement between an input and its transformations while minimizing agreement between transformations of an input. The same deterministic loss function is used as the scoring function.

\subsection{Results and Discussion}
\begin{table*}[ht]
\caption{Average Precision, Recall, and F1-Score (all with standard deviation) of the six models trained exclusively on normal samples contaminated with a ratio $c$ of attacks samples.}
\label{f1_results_table}
\vskip 0.15in
\centering
\tiny
\begin{tabular}
{@{}lclllllllllllll@{}}
\toprule & & \multicolumn{3}{c}{KDDCUP 10} & \phantom{a} & \multicolumn{3}{c}{NSL-KDD} & \phantom{a} & \multicolumn{3}{c}{CSE-CIC-IDS2018} \\
\cmidrule{3-5} 
\cmidrule{7-9} 
\cmidrule{11-13
} 
& Ratio $c$ & Precision  & Recall & $F_1$ && Precision & Recall & $F_1$ && Precision & Recall & $F_1$  \\
\hline

 \multirow{4}{*}{ALAD} & 0\% & 92.5$\pm$0.9 & 97.2$\pm$5.5 & 94.7$\pm$3.2 &     & 87.1$\pm$2.5 & 91.9$\pm$1.1 & 89.4$\pm$1.4 &      & 47.6$\pm$8.4 & 52.0$\pm$5.8 & 49.5$\pm$6.6\\
                       & 5\% & 90.6$\pm$4.8 & 98.5$\pm$1.0 & 94.3$\pm$3.0 &     & 88.2$\pm$2.3 & 88.2$\pm$4.2 & 88.2$\pm$2.8 &      & 36.9$\pm$2.4 & 43.4$\pm$5.4 & 39.8$\pm$3.6\\
                       & 8\% & 80.5$\pm$9.6 & 90.6$\pm$15.8 & 85.0$\pm$12.4 &     & 87.3$\pm$3.9 & 87.5$\pm$2.2 & 87.3$\pm$2.0 &      & 34.9$\pm$7.1 & 36.8$\pm$6.2 & 35.7$\pm$6.1\\
                       & 12\% & 77.6$\pm$9.6 & 89.0$\pm$17.3 & 82.6$\pm$13.2 &    & 80.6$\pm$4.3 & 87.0$\pm$4.5 & 83.7$\pm$4.4 &      & 33.0$\pm$4.0 & 37.5$\pm$4.4 & 35.0$\pm$3.7\\

\cmidrule{1-13}

\multirow{4}{*}{DAE} & 0\% & 88.8$\pm$0.4 & 99.4$\pm$0.4 & 93.8$\pm$0.1 && 92.9$\pm$1.1 & 95.4$\pm$1.2 & 94.1$\pm$0.1 && 69.5$\pm$3.6 & 57.6$\pm$0.8 & 63.0$\pm$1.5\\
                       & 5\% & 79.6$\pm$7.5 & 86.4$\pm$10.2 & 82.7$\pm$8.3 && 82.1$\pm$2.5 & 86.4$\pm$4.0 & 84.1$\pm$1.7 && 54.9$\pm$5.2 & 52.0$\pm$3.8 & 53.1$\pm$2.2\\
                       & 8\% & 68.5$\pm$3.8 & 75.1$\pm$10.2 & 71.4$\pm$5.9 && 83.2$\pm$2.4 & 80.6$\pm$2.6 & 81.8$\pm$0.5 && 43.9$\pm$2.9 & 47.2$\pm$4.0 & 45.3$\pm$2.5\\
                       & 12\% & 69.0$\pm$7.2 & 71.4$\pm$8.5 & 69.9$\pm$6.2 && 79.7$\pm$2.8 & 82.6$\pm$2.5 & 81.1$\pm$1.0 && 43.2$\pm$3.8 & 41.6$\pm$5.3 & 42.0$\pm$2.2\\

\cmidrule{1-13}

\multirow{4}{*}{DAGMM} & 0\% & 91.2$\pm$4.9 & 96.7$\pm$3.0 & 93.8$\pm$3.6    && 86.4$\pm$4.2 & 85.9$\pm$2.4 & 86.0$\pm$1.8 && 42.4$\pm$11.6 & 41.3$\pm$10.1 & 41.7$\pm$10.8\\
                       & 5\% & 76.1$\pm$21.0 & 77.3$\pm$21.6 & 76.3$\pm$20.9 && 83.0$\pm$6.0 & 86.5$\pm$4.6 & 84.7$\pm$5.1 && 31.6$\pm$14.2 & 29.9$\pm$8.4 & 30.4$\pm$11.0\\
                       & 8\% & 60.4$\pm$11.7 & 59.0$\pm$16.8 & 59.1$\pm$13.4 && 77.9$\pm$9.5 & 81.4$\pm$7.8 & 79.5$\pm$8.3 && 23.7$\pm$10.1 & 28.3$\pm$12.9 & 25.7$\pm$11.3\\
                       & 12\% & 42.8$\pm$13.5 & 39.3$\pm$14.9 & 40.7$\pm$13.9 && 75.9$\pm$14.0 & 79.7$\pm$13.5 & 77.7$\pm$13.6 && 13.4$\pm$7.8 & 20.6$\pm$12.7 & 15.9$\pm$9.0\\

\cmidrule{1-13}

\multirow{4}{*}{DSEBM} & 0\% & 94.1$\pm$0.3 & 99.4$\pm$0.1 & 96.7$\pm$0.1  && 93.6$\pm$0.3 & 93.1$\pm$0.4 & 93.4$\pm$0.1 && 29.6$\pm$0.1 & 37.0$\pm$0.4 & 32.9$\pm$0.2\\
                       & 5\% & 88.1$\pm$0.1 & 100.0$\pm$0.0 & 93.7$\pm$0.0 && 86.0$\pm$1.6 & 93.2$\pm$1.8 & 89.5$\pm$1.7 && 29.2$\pm$0.1 & 36.6$\pm$0.1 & 32.5$\pm$0.1\\
                       & 8\% & 87.3$\pm$0.2 & 99.6$\pm$0.2 & 93.1$\pm$0.2  && 77.3$\pm$0.7 & 84.0$\pm$0.8 & 80.5$\pm$0.7 && 29.0$\pm$0.1 & 36.5$\pm$0.2 & 32.3$\pm$0.1\\
                       & 12\% & 86.7$\pm$0.4 & 99.1$\pm$0.3 & 92.5$\pm$0.4 && 77.5$\pm$0.4 & 79.5$\pm$0.2 & 78.5$\pm$0.2 && 28.6$\pm$0.1 & 35.7$\pm$0.2 & 31.8$\pm$0.2\\

\cmidrule{1-13}

\multirow{4}{*}{DUAD} & 0\% & 94.0$\pm$0.5 & 99.4$\pm$0.2 & 96.6$\pm$0.2   && 91.9$\pm$0.7 & 91.6$\pm$0.5 & 91.7$\pm$0.1 && 62.2$\pm$3.9 & 73.8$\pm$0.6 & 67.5$\pm$2.3\\
                       & 5\% & 88.0$\pm$0.2 & 99.9$\pm$0.1 & 93.6$\pm$0.1  && 92.5$\pm$0.9 & 90.9$\pm$1.1 & 91.7$\pm$0.2 && 51.9$\pm$5.2 & 62.1$\pm$5.9 & 56.5$\pm$5.4\\
                       & 8\% & 85.0$\pm$1.8 & 97.8$\pm$2.2 & 90.9$\pm$1.5  && 92.9$\pm$1.3 & 90.3$\pm$1.6 & 91.6$\pm$0.2 && 46.7$\pm$6.8 & 56.9$\pm$8.0 & 51.3$\pm$7.2\\
                       & 12\% & 86.1$\pm$0.3 & 90.0$\pm$0.8 & 88.0$\pm$0.5 && 90.8$\pm$1.0 & 92.2$\pm$0.6 & 91.5$\pm$0.4 && 41.2$\pm$1.9 & 46.9$\pm$7.2 & 43.7$\pm$4.1\\

\cmidrule{1-13}

\multirow{4}{*}{NeuTraLAD} & 0\% & 88.2$\pm$0.4 & 99.9$\pm$0.1 & 93.7$\pm$0.2 && 95.0$\pm$0.9 & 93.8$\pm$0.7 & 94.4$\pm$0.3 && 69.2$\pm$4.0 & 55.7$\pm$0.9 & 61.7$\pm$1.4\\
                           & 5\% & 85.1$\pm$0.5 & 98.0$\pm$1.4 & 91.1$\pm$0.7 && 87.0$\pm$1.8 & 94.3$\pm$1.9 & 90.5$\pm$1.9 && 47.2$\pm$5.4 & 51.9$\pm$7.6 & 49.1$\pm$5.5\\
                           & 8\% & 81.5$\pm$2.7 & 94.0$\pm$2.2 & 87.2$\pm$2.1 && 82.5$\pm$1.7 & 88.1$\pm$2.7 & 85.2$\pm$1.6 && 27.3$\pm$13.2 & 33.3$\pm$16.2 & 30.0$\pm$14.5\\
                           & 12\% & 75.4$\pm$3.2 & 89.4$\pm$2.8 & 81.7$\pm$2.5 && 81.2$\pm$2.7 & 83.2$\pm$3.0 & 82.1$\pm$0.9 && 24.2$\pm$11.8 & 25.8$\pm$14.2 & 24.8$\pm$12.8\\

\bottomrule
\end{tabular}
\end{table*}

Table \ref{f1_results_table} reports F1-score, precision, and recall of the six models with different values of contamination ratio $c$. 

\textbf{Performance on KDDCUP dataset.} With no contamination, all six models perform well, with an average F1-score above 90\% over the 20 runs; DSEBM and DUAD outperform other models with an F1-score of 96\%. From the two-dimensional representation of a subset of the KDDCUP dataset in Figure \ref{fig:tsnekdd}, it can be seen that the normal and attacks data are mostly distinguishable -- we observe a less proportion of overlapping between the two. This could explain why all models perform well on KDDCUP dataset. From Figure \ref{fig:f1kdd}, we notice a loss of performance, albeit in different proportions,  for all six models as the contamination ratio increases. Remarkably, the F1-scores of DAGMM and DAE drop dramatically from 93.8\% to 40.7\% and from 93.8\% to 69.9\%, respectively, with a contamination ratio of 12\%. 

DAGMM estimates a gaussian mixture with no explicit way to handle outliers. Consequently, contaminated samples influence density estimation and move normal samples towards low-density regions. However, DUAD and DSEBM have demonstrated some resistance to contamination. DUAD's re-evaluation trick on the training data appears to have discarded some contaminated samples to allow training on a nearly clean subset.

As for DSEBM, we note that the recall remains at approximately 99\%, but only the precision decreases. That means DSEBM was able to classify all attacks well, assigning high energy to them, regardless of the contamination level. However, the precision of DSEBM decreases as the noise in the training data increases. 

\textbf{Performance on NSL-KDD dataset.} Figure \ref{fig:f1nsl} shows different patterns than those observed on the KDDCUP dataset. Note that NSL-KDD is a relatively balanced dataset. For example, DSEBM yields poor performance with an F1-score that drops from 93\% to 78\% for a contamination rate of 12\%. That may be due to the removal of duplicates in the KDDCUP dataset. On the other hand, DUAD distribution clustering plays a central role of defending against contamination of the training set. We observe that DUAD consistently performs well (F1 score of 91\%), regardless of the contamination level.

\textbf{Performance on CSE-CIC-IDS2018 dataset.} On this dataset, DUAD achieves the highest F1-score (67\%), followed by the DAE (63\%) and NeuTraLAD  (61\%). Overall, all six models struggle to achieve high performance with regards to what they yielded on KDDCUP and NSL-KDD datasets, even without contamination. 

In Figure \ref{fig:tsnecic}, we visualize, by t-SNE, a subset of normal and infiltration attack data from CSE-CIC-IDS2018 dataset. Surprisingly, we observe an important number of overlaps between the benign traffic and the attack data, so this makes it difficult to distinguish between them. Figure \ref{fig:f1cic} shows an almost linear decrease in F1-scores for all six models as the contamination rate increases. DUAD has previously demonstrated resistance to contamination, but its F1 score drops from 67\% to 43\% on CSE-CIC-IDS2018 with a contamination rate of 12\%. This poor performance is because, unlike KDDCUP and NSL-KDD (Figures~\ref{fig:tsnekdd}~and~\ref{fig:tsnensl}), some of the attack samples in CSE-CIC-IDS2018 (Figure \ref{fig:tsnecic}) are less different from normal traffic data. Therefore, it is difficult to eliminate contamination from the training set, even for DUAD. Nevertheless, DUAD still obtains a higher F1-score than the other models at different contamination rates; followed by DAE and then by ALAD.  

\begin{figure}[ht]{}
         \centering
         \includegraphics[width=\columnwidth]{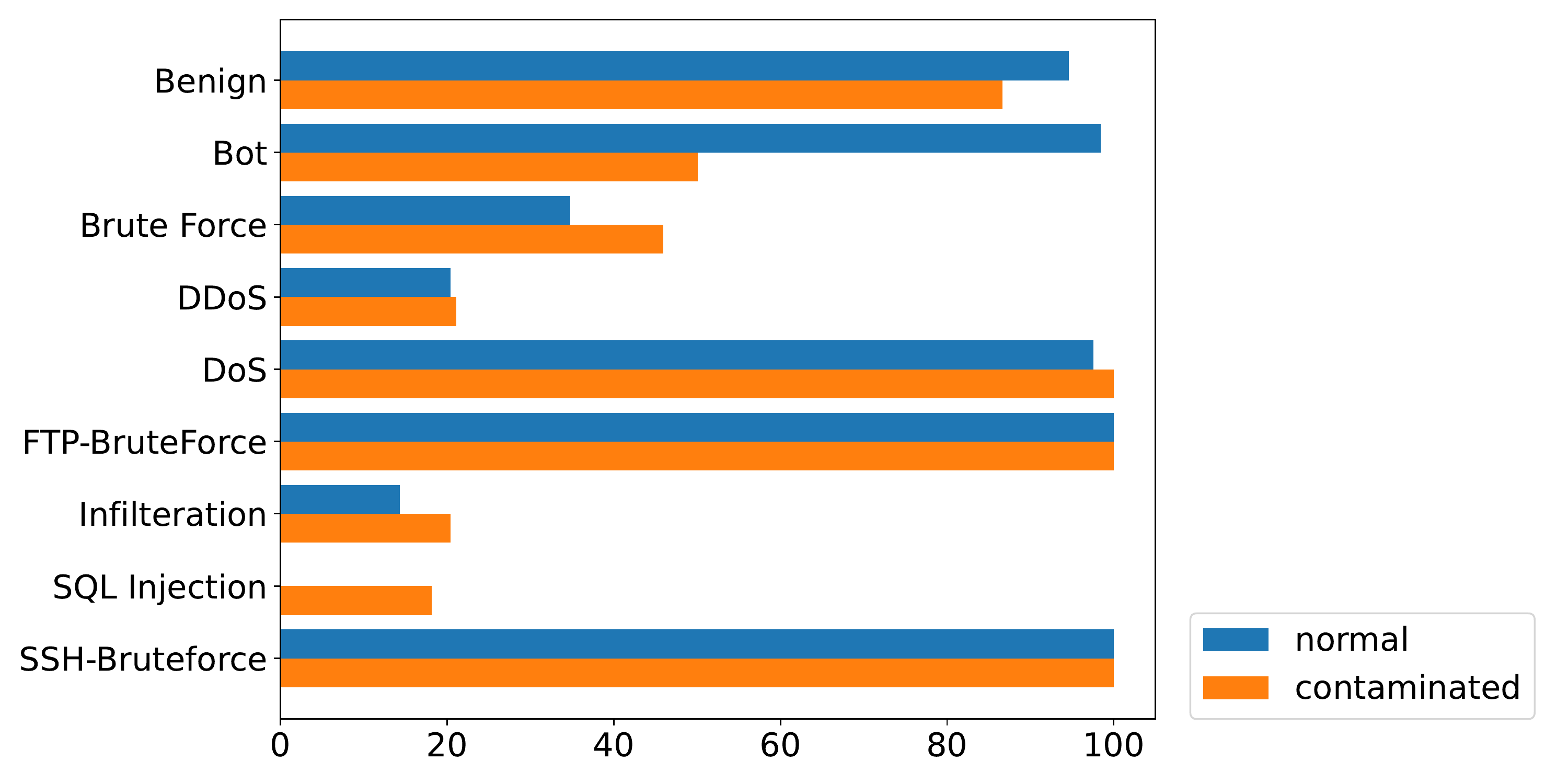}
         \caption{DAE accuracy on CSE-CIC-IDS2018 dataset grouped by class, when trained exclusively on normal data and when trained on normal data plus a ratio of five percent of randomly selected attack data.}
        \label{fig:accuracybyclass05}
        \vskip 0.2in
\end{figure}


\begin{figure}[ht]{}
\vskip 0.2in
         \centering
         \includegraphics[width=\linewidth]{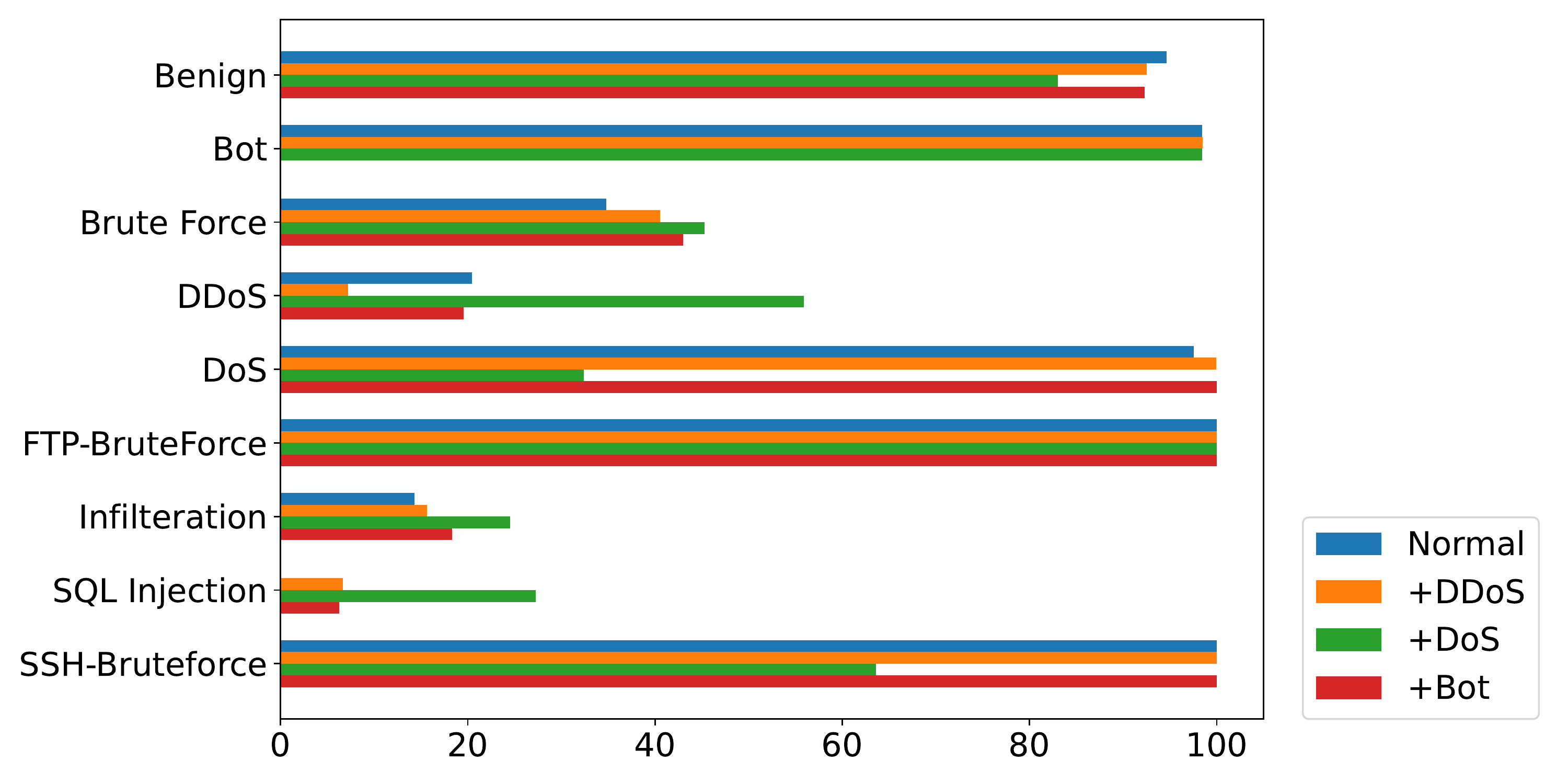}
        \caption{DAE accuracy on CSE-CIC-IDS2018 dataset -- grouped by class (benign and types of attacks) -- when trained on normal samples only, compared to when training is performed on normal samples mixed with a ratio of one percent of DDoS, DoS, and Bot attack, exclusively.}
        \label{fig:accuracybyclassdosddosbot}
\end{figure}

\textbf{Accuracy per class on CSE-CIC-IDS2018 dataset.} We present the analysis of DAE detection accuracy at data class (benign and attacks type) level. Figure \ref{fig:accuracybyclass05} shows accuracy for each type of attack and benign traffic when DAE is solely trained on normal data or on normal data mixed with 5\% of attack data, randomly drawn from the contamination set. When training is performed on normal data, we notice that some attacks, such as DDoS, infiltration, and SQL Injection, have a low detection rate, below 40\%. One cause of these inaccurate predictions may be the similarity between some normal traffic and attack data as displayed in Figure \ref{fig:tsnecic}.
Training set contamination modifies the decision boundary. Still, in Figure \ref{fig:accuracybyclass05}, it can be observed that if the training set is at 5\% corrupted, the detection rate of benign traffic decreases, while the detection rate of some attacks, including infiltration, and SQL Injection attacks, increases. On the one hand, the number of false positive samples increases, and on the other hand, some attacks are now being recognized by the model, although their accuracies remain low. Yet, the overall performance is lower, due to the imbalanced nature of classes in the dataset.

Instead of contaminating the training set with all types of attacks, we found that it is important to investigate the impact of corrupting the training set with specific types of attacks. Figure \ref{fig:accuracybyclassdosddosbot} shows the accuracy of DAE when trained only on normal data and when trained on normal data contaminated by a ratio of 1\% of DDoS attacks, DoS attacks, and Bot attacks, exclusively. We observe that contamination with either DDoS, DoS, or Bot drops F1-scores from 63\% to 57\%, 50\%, and 53\%, respectively. It is important to note that with only 1\% of the training set contamination with the Bot attack, F1 score drops and the Bot attack is absolutely undetected by DAE. That is an example of a backdoor attack scenario where an attacker pollutes the training set to allow a subsequent undetected intrusion at runtime.

\textbf{Summary.} Based on our experiments, we see that the contamination of the training set can negatively impact the performance of even the most advanced models. Since there is no guarantee that data will always be clean, it is therefore critical to implement a defense against contamination when developing ML models for cybersecurity. One way of defense could be to infer data labels during training steps. Once inferred labels are available, the algorithm would eject from the training dataset samples likely to be anomalous.

We note that the KDDCUP and NSL-KDD datasets contain attacks far from reflective of current cyber-attacks. Nowadays, hackers use advanced tools, including AI-based ones, to generate sophisticated attacks. These types of attacks remain challenging and difficult to detect. KDDCUP and NSL-KDD datasets are no longer suitable for simulating real computer networks.  The results obtained on KDDCUP and NSL-KDD could be misleading compared to actual reality. For example, models with superior performance on these two datasets do not perform as well on the CSE-CIC-IDS2018 dataset. The latter seems challenging and relatively suitable for simulating the current state of computer networks and cyber-attacks.


We observe that DUAD consistently showed some resistance to contamination, thanks to its re-evaluation of the training set through clustering. One drawback of DUAD is that clustering is performed over the entire training set, once in data space, then in latent space repeatedly after a fixed number of epochs until convergence. This renders DUAD very slow to train for large datasets. Alternatively, we could extend the idea of distribution clustering (introduced by DUAD) of the training set to reject contaminated data in an online fashion.







\section{Conclusion}
\label{conclu}
In this paper, we evaluate the robustness of state-of-the-art anomaly detection models on network intrusion detection datasets with different levels of training set contamination. We show that model performance drops when the training set is poisoned by attack instances, even for modern deep learning models. Furthermore, our study reveals the importance of robustness to contamination as a criterion when choosing an anomaly detection model for cybersecurity applications. We also highlight how model performance on outdated cybersecurity datasets could be misleading with respect to the current state of computer networks and types of attacks. We aim to develop an appropriate defense framework against training set contamination for deep anomaly detection methods applied to cybersecurity as complementary work.

\section*{Acknowledgements}
We acknowledge the support of Hydro-Sherbrooke, Natural Resources Canada (NRCan) and Public Safety Canada for this work.

\bibliography{example_paper}
\bibliographystyle{icml2022}

\end{document}